\documentclass{osa-article}
\usepackage{soul}
\journal{osajournal}
\usepackage{float}
\usepackage{mathtools}
\usepackage[mathscr]{eucal}
\articletype{Research Article}
\usepackage[euler]{textgreek}
\usepackage{changes}

\begin{document}

\title{Full 3D+1 modelling of tilted-pulse-front setups for single-cycle terahertz generation}

\author{Lu Wang,\authormark{1,2*} Tobias Kroh,\authormark{1,2} Nicholas H. Matlis\authormark{1} and Franz K\"{a}rtner\authormark{1,2,3}}

\address{\authormark{1}Center for Free Electron Laser Science (CFEL), Deutsches Elektronen-Synchrotron, Hamburg, 22607, Germany\\
\authormark{2}Universit\"{a}t of Hamburg, Department of Physics, Hamburg,  22761, Germany\\
\authormark{3}The Hamburg Centre for Ultrafast Imaging (CUI), Hamburg, 22761, Germany}

\email{\authormark{*}lu.wang@desy.de} 



\begin{abstract}
The tilted-pulse-front setup utilizing a diffraction grating is one of the most successful methods to generate single- to few-cycle terahertz pulses. However, the generated terahertz pulses have a large spatial inhomogeneity, due to the noncollinear phase matching condition and the asymmetry of the prism-shaped nonlinear crystal geometry, especially when pushing for high optical-to-terahertz conversion efficiency. A 3D+1 (x,y,z,t) numerical model is necessary in order to fully investigate the terahertz generation problem in the tilted-pulse-front scheme. We compare in detail the differences between 1D+1, 2D+1 and 3D+1 models. The simulations show that the size of the optical beam in the pulse-front-tilt plane sensitively affects the spatio-temporal properties of the terahertz electric field. The terahertz electric field is found to have a strong spatial dependence such that a few-cycle pulse is only generated near the apex of the prism. Even though the part of the beam farther from the apex can contain a large fraction of the energy, the terahertz waveform shows less few-cycle character. This strong spatial dependence must be accounted for when using the terahertz pulses for strong-field physics and carrier-envelope-phase sensitive experiments such as terahertz acceleration, coherent control of antiferromagnetic spin waves and terahertz high-harmonic generation. 
\end{abstract}

\section{Introduction}

Single- to few-cycle (broadband) high energy terahertz pulses have many promising applications such as spectroscopy \cite{davies2002development}, strong field terahertz physics\cite{kampfrath2011coherent,schubert2014sub}, particle acceleration \cite{zhang2018segmented} , electron spin manipulation \cite{kampfrath2011coherent} and phonon resonance studies \cite{bakker1992observation}. All of these applications require well-characterized terahertz fields. 

 There are many possible ways to generate terahertz radiation. Free electron lasers and synchrotron radiation have a high degree of tunability and are capable of delivering high peak-power coherent terahertz pulses \cite{tan2012terahertz}. Gyrotrons, based on the principle of electron cyclotron radiation, are able to generate watt-to-megawatt-level terahertz continuous wave radiation \cite{glyavin2008generation} at low terahertz frequencies (0.3-1.3 THz)\cite{bratman2009large,bratman2012gyrotron}. These devices, however, have limited accessibility to the larger scientific community, and can be difficult to synchronize to laser sources with high (fs) precision. 
 
Alternatively, single- to few-cycle terahertz generation, based on table-top optical laser systems, brings the advantages of high accessibility and intrinsic synchronization, but suffers from limited optical-to-terahertz conversion efficiency. 
Although various schemes exist for optical-to-terahertz conversion, difference frequency generation using the tilted pulse-front (TPF) method has proven one of the most useful. Using the "tilted-pulse-front" technique, first proposed and demonstrated by J.Hebling et.al.\cite{hebling2002velocity}, pulse energies in the milijoule range can be reached\cite{fulop2014efficient}. The ease of this setup, the high pulse energies and the controllability of the terahertz properties has made this last approach an ubiquitous one for high-field applications.

However, the non-collinear geometry of the phase matching and the spatial asymmetry of the interaction, in combination with the cascading effect, result in terahertz beams with non-uniform spatial distribution. This non-uniformity was studied by M. I. Bakunov et al. \cite{bakunov2008terahertz,bakunov2011terahertz} via a 2D+1 numerical model. However, this model doesn't include the back conversion of the terahertz to the optical pump (OP),i.e the cascading effect. Consequently, the effective length and the conversion efficiency are overestimated. Later on, the interaction between the optical pump and the terahertz pulse was included into the 2D+1 model along with a one lens imaging system by K. Ravi et al. \cite{ravi2015theory}. The one-lens imaging system, has larger imaging errors and induces more terahertz divergence \cite{tokodi2017optimization}, compared with the telescope imaging system. In order to fully investigate the non-uniform spatial distribution of the generated terahertz pulse, a robust 3D+1 numerical tool, including cascading effect and the telescope imaging system, is necessary. Additionally,we systematically compare 1D+1, 2D+1 and 3D+1 numerical models regarding their practical relevance for the first time to our knowledge.

The numerical tool is based on the fast Fourier transform beam propagation method (FFT-BPM) \cite{feit1978light} and split-step Fourier method. The combination of these two methods reduces computational cost compared to the finite difference time-domain (FDTD) method which is very accurate but  requires a massive computational effort. In section 2, we show analytically that the higher order dispersion introduced by the grating can cause spatial inhomogeneity of the generated terahertz pulses. In section 3, the numerical model including the nonlinear interactions which lead to further spatial inhomogeneity is introduced. In addition, the differences of the 1D+1, 2D+1 and 3D+1 models are presented. Finally, in section 4, the dependence of optical pump beam sizes on the generated terahertz fields are discussed.

\section{Theoretical model}
\begin{figure}[H]
\centering{
   \includegraphics[width=0.7\textwidth]{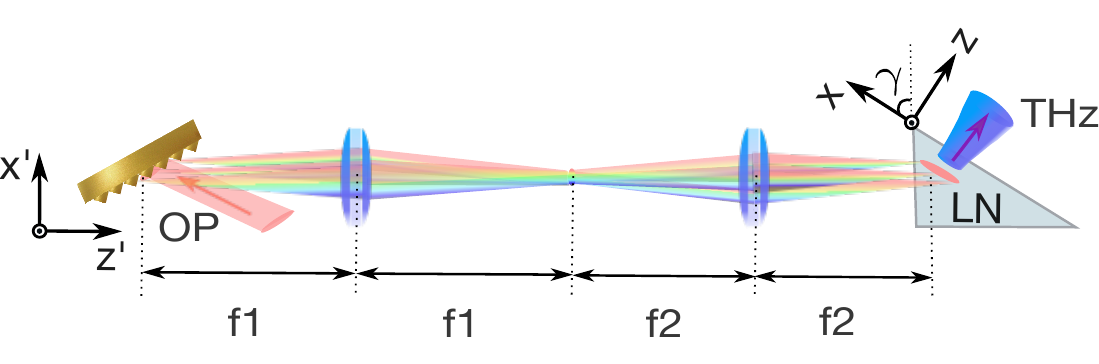}}
   \caption{Illustration of the simulated tilted-pulse-front setup. The optical pump pulse is noted by OP and the $\text{LiNbO}_3$ crystal is represented by LN. The OP propagates along the $z'$ direction. The $x-y-z$ coordinates denote the pulse-front-tilt frame (terahertz frame) inside the LN-crystal. The $y$ and $y'$ axes are equivalent. The optical system is chosen such that the image of the grating is parallel to the pulse-front-tilt inside the LN\cite{fulop2010design} . }\label{setup}
\end{figure}

The tilted-pulse-front setup modelled and simulated is shown in Fig. \ref{setup}. Often used nonlinear materials are $\text{LiNbO}_3$ CdTe, GaAs, GaSe, GaP and ZnTe. Here, we focus on $\text{LiNbO}_3$ (LN) due to its large second-order nonlinear coefficient, high damage threshold and easy accessibility. The results however can be extended to other materials. In this article, we focus on the impact of the OP beam size on the spatio-temporal properties of the generated terahertz beam.

First, we show analytically that, regardless of the noncollinear phase matching and the prism geometry, the second order dispersion generated by the grating can already cause inhomogeneity of the terahertz pulses (see Eq. (\ref{polari})).   We start from the grating relation $\sin(\theta_1)+\sin(\theta_2)=2\pi c/\omega d$, where $\theta_1$, $\theta_2$ are the incidence and output angles with respect to the normal to the grating surface respectively. The optical frequency is represented by $\omega$ while $d$ denotes the grating groove period. Assuming that the input OP is collimated ($d \theta_1/d \omega =0$), one can get Eq. (\ref{grating_1}).

\begin{equation}
\begin{dcases}\label{grating_1}
    \frac{d\theta_2}{d \omega}=\frac{-2\pi c}{\omega^2d\cos(\theta_2)}\\
        \frac{d^2\theta_2}{d^2 \omega}=\frac{4\pi c}{\omega^3d\cos(\theta_2)}-\frac{2\pi c\sin(\theta_2)}{\omega^2d\cos^2(\theta_2)}\frac{d\theta_2}{d\omega}
\end{dcases}
\end{equation}

One can obtain the angular dispersion $\Delta \theta_2$ with respect to the OP propagation direction, via performing a Taylor expansion to the second order. By inserting Eq. (\ref{grating_1}) and setting the first order angular dispersion $F_1=-2\pi c/(\omega_0^2 d \cos(\theta_{02}))$ where $\omega_0$ is the center frequency and $\theta_{02}$ is the grating output angle at the center frequency, we obtain:

\begin{eqnarray}\label{grating_a_d}
\Delta \theta_2&&=\frac{d \theta_2}{d \omega}|_{\omega=\omega_0}(\omega-\omega_0)+\frac{1}{2}\frac{d^2 \theta_2}{d^2 \omega}|_{\omega=\omega_0}(\omega-\omega_0)^2 \nonumber \\
&&=F_1\,(\omega-\omega_0)+\left[-F_1/\omega_0+\frac{1}{2}F_1^2\tan(\theta_{02})\right](\omega-\omega_0)^2 \nonumber \\
&&=F_1\,(\omega-\omega_0)+F_2(\omega-\omega_0)^2
\end{eqnarray}
The second order dispersion is denoted by $F_2$. After the propagation through the telescope system, the angular dispersion is magnified by a factor of -$f_1/f_2$. Thus, the angular dispersion becomes $-\Delta\theta_2 f_1/f_2$. Note that the transverse k-vector $k_{x'0}(\omega) = -\Delta\theta_2 f_1\omega_0/(cf_2)$ remains the same before and after entering the LN-crystal due to the Fresnel law.  As a result, by assuming $ k_{z'0}(\omega) \approx \omega n(\omega)/c$, where $n$ is the refractive index, the OP electric field inside the LN-crystal is given in Eq. (\ref{input}) in the $x'-y'-z'$ coordinates. In Eq. (\ref{input}), the pulse duration is $\tau = \tau_0/\sqrt{2\log{2}}$ , $A_0$ is the electric field amplitude and $\sigma_x',\sigma_y'$ are beam waists (1/e) of the OP at the incidence surface of the LN-crystal in $x'$ and $y'$ dimensions respectively.

\begin{eqnarray}\label{input}
E(\omega,x',y',z')=&&A_0\exp{\left[-(\omega-\omega_0)^2\tau^2/4\right]}\exp{\left[-x'^2/(2\sigma_x'^2)\right]}\exp{\left[-y'^2/(2\sigma_y'^2)\right]} \nonumber\\
&&\times \exp{\Bigl[-i\omega n(\omega)z'/c \Bigr]}\exp{\Bigl[i\Delta\theta_2 f_1\omega_0x'/(cf_2)\Bigr]}
\end{eqnarray}
The imaging system was chosen such that the image of the grating is parallel to the pulse-front-tilt plane inside the nonlinear crystal \cite{fulop2010design}. The expression of the OP electric field given in Eq. (\ref{input}), is only valid at the imaging plane of the telescope. It is very important to notice that the higher order angular dispersion ($F_2$) induced by the grating leads to temporal broadening of the OP\cite{martinez1988matrix,martinez1986grating}, as it adds nonlinear phase (see Eqs. (\ref{grating_a_d}) and (\ref{input})). If the second order angular dispersion is neglected ($F_2=0$ in Eq. (\ref{grating_a_d})), the OP pulse duration reduces to the transform limited case at each spatial point along the $x'$ dimension. 

Using Eq. (\ref{input}), the second order polarization which is responsible for the terahertz generation can be expressed as in Eq. (\ref{polari}), where $\gamma$ is the angle between the terahertz pulse propagation direction and the OP propagation direction, $\chi^{(2)}$ is the second order nonlinear susceptibility and $\Omega$ is the terahertz angular frequency. The group refractive index at the center frequency of the optical pump is denoted by $n_g$.
\begin{eqnarray}\label{polari}
P^{(2)}_{NL}(\Omega,x',y',z')=&&-\chi^{(2)}\frac{\Omega^2}{c^2}\int_{0}^{\infty}E(\omega+\Omega,x',y',z')E^*(\omega,x',y',z')d\omega  \nonumber \\
&&\times \exp{\left\{i\Omega n(\Omega)[\cos(\gamma)z'+\sin(\gamma)x']/c \right\} } \nonumber \\
=&&-\chi^{(2)}\frac{\Omega^2\sqrt{2\pi}}{\tau c^2}A_0^2\exp{\left(\frac{-x'^2}{\sigma_x'^2}\right)}\exp{\left(\frac{-y'^2}{\sigma_y'^2}\right)}\nonumber \\
&&\times \exp{\left(-\frac{\Omega^2\tau^2}{8}\left\{1+\frac{16x'^2n_g^2\tan(\gamma)^2}{\tau^4c^2}\left(\frac{F_2}{F_1}\right)^2\right\}\right)} \nonumber\\
&&\times \exp\left\{ -i\frac{\Omega }{c}\Bigl[ n_g-n(\Omega)\cos(\gamma)\Bigr] z'\right\}\exp\left\{ i\Bigl[\frac{f_1\omega_0}{f_2}F_1+n(\Omega)\sin(\gamma)\Bigr]\frac{\Omega}{c}x'\right\}\nonumber \\
\end{eqnarray}

The last two exponential phase terms in Eq. (\ref{polari}) represent the phase matching condition.
\[
\begin{dcases}
\Delta k_z'=\frac{\Omega }{c}\Bigl[ n_g-n(\Omega)\cos(\gamma)\Bigr]=0 \rightarrow n_g/n(\Omega)=\cos(\gamma)\\
\Delta k_x'=\Bigl[\frac{f_1\omega_0}{f_2}F_1+n(\Omega)\sin(\gamma)\Bigr]\frac{\Omega}{c}=0 \rightarrow  \frac{2\pi cf_1}{d\cos(\theta_2)\omega_0 n_gf_2}=\tan(\gamma)
\end{dcases}
\]

The term in the third exponential ($\left(F_2/F_1\right)^2{16x'^2n_g^2\tan(\gamma)^2}/{\tau^4c^2}$ ) is due to the second order angular dispersion which has been discussed in more detail in  \cite{ravi2019analysis}. It can be seen that the second order angular dispersion leads to a spatial dependence of the generated terahertz bandwidth along the $x'$ dimension. At the center of the pump pulse ($x'=0$), the generated terahertz pulse possesses its largest bandwidth. However, towards the sides of the OP beam, the bandwidth of the terahertz pulse reduces. In other words, due to the second order angular dispersion ($(\omega-\omega_0)^2$ related term), the OP experiences a temporal chirp and thus, the pulse duration varies with respect to $x'$. The effective instantaneous bandwidth of the OP reduces towards the sides of the beam, leading to a narrower terahertz spectrum (multi-cycle pulses).

\section{Comparison of the 1D+1, 2D+1 and 3D+1 simulations}
Owing to the trapezoid geometry of the LN-crystal, the interaction length varies with respect to $x'$. Since the generated terahertz pulses act back onto the OP, cascading occurs. This causes a spatially dependent OP spectrum, which further enhances spatial inhomogeneities of the generated terahertz pulses. Furthermore, at the desired terahertz frequency range (<4 THz), the material absorption ($\alpha$) increases with respect to frequency. This favors lower terahertz frequencies towards the base of the LN-crystal due to a longer interaction length. The aforementioned effects contribute to the spatial inhomogeneity of the terahertz pulses, in addition to the higher order angular dispersion caused by the grating (Section. 2). The combined effect can only be investigated by a robust numerical model.

Our numerical model solves the coupled wave equations with slowly varying amplitude approximation in the terahertz coordinates (x-y-z). By setting the electric field of the OP $\mathscr{F}{[E_\text{op}(t,x,y,z)]}=E_\text{op}(\omega,x,y,z)=E(\omega,x,y,z)e^{-i[k_{z0}(\omega)z+k_{x0}(\omega)x]}$, and the electric field of the terahertz
$\mathscr{F}{[E_\text{THz}(t,x,y,z)]}=E_\text{THz}(\Omega,x,y,z)=E(\Omega,x,y,z)e^{-ik_{0}(\Omega)z}$ respectively,  one can get Eqs. (\ref{eq1}) and (\ref{eq2}). The operator $\mathscr{F}$  represents the Fourier transform.
 \begin{eqnarray}
-2ik_0(\Omega)\frac{\partial E(\Omega,x,y,z)}{\partial z}=-[\underbrace{\overbrace{\frac{\partial^2}{\partial y^2}}^{\mathclap{\text{neglected by 2D+1}}}+\frac{\partial^2}{\partial x^2}}_\text{neglected by 1D+1}-i\alpha k_{0}(\Omega)]E(\Omega,x,y,z)&& \nonumber \\
-\frac{\Omega^2 \chi^{(2)}}{ c^2 }\int_{0}^\infty E(\omega+\Omega,x,y,z)E^*(\omega,x,y,z)e^{i(\Delta k_zz+\Delta k_xx)}d\omega&& \label{eq1}\\
\nonumber \\
-2ik_{z0}(\omega)\frac{\partial E(\omega,
x,y,z)}{\partial z}=-[\underbrace{\overbrace{\frac{\partial^2}{\partial y^2}}^{\mathclap{\text{neglected by 2D+1}}}+ \frac{\partial^2}{\partial x^2} -2ik_{x0}(\omega)\frac{\partial}{\partial x} }_\text{neglected by 1D+1}]  E(\omega,x,y,z) &&\nonumber \\
 -\varepsilon_0 n^2(\omega_0) \frac{\omega^2}{c}\mathscr{F}{\left\{ E_{\mathrm{op}}(t,x,y,z)\int_{-\infty}^{\infty} n_2(\tau)E^2_{\mathrm{op}}(t-\tau,x,y,z)d\tau \right\}}e^{i[k_{z0}(\omega)z+k_{x0}(\omega)x]} &&\nonumber \\
-\frac{\omega^2 \chi^{(2)}}{ c^2 }\int_{-\infty}^\infty E(\omega+\Omega,x,y,z)E^*(\Omega,x,y,z)e^{i(\Delta k_zz+\Delta k_xx)}d\Omega && \label{eq2}
 \end{eqnarray}
 In Eqs. (\ref{eq1}) and (\ref{eq2}), $\Delta k_x=k_{x0}(\omega)-k_{x0}(\omega+\Omega)$, $\Delta k_z=k_{z0}(\omega)-k_{z0}(\omega+\Omega)+k_{0}(\Omega)$. The $\chi^{(2)}$ related terms are responsible for the second order nonlinear effects, i.e., the terahertz generation and back conversion processes. In Eq. (\ref{eq2}), the third-order nonlinear effects, including self-phase-modulation, self-steepening and stimulated Raman effect, are represented by the term $n_2(\tau)=\mathscr{F}{[n_2(\omega-\omega_0)]}$ \cite{hellwarth1977third}, where $n_2$ is the nonlinear refractive index.  The phonon resonances at terahertz frequencies \cite{sussman1970tunable} are implemented by  considering the stimulated Raman effect at the optical frequency region together with the frequency-dependent refractive index in the terahertz frequency region. In the simulation, the refractive index and terahertz absorption are frequency dependent. The parameters used are listed in Table \ref{tpf_parameters}. The peak fluence of the OP at the input LN-crystal surface is chosen to be right beneath the estimated damage threshold $70.7$\,mJ/cm$^2$ based on our previous studies\cite{ravi2016pulse}.
\begin{table}[H]
\caption {Simulation parameters} \label{tpf_parameters} 
\begin{center}
    \begin{tabular}{||p{2.5cm}|p{1.5cm}||p{3.1cm}|p{3cm}||}
    \hline
   \textbf{Parameters} & \textbf{Value} & \textbf{Parameters} & \textbf{Value} \\ \hline
    focal length ${f}_1$  & 300\,mm  & focal length ${f}_2$  & 0.613 $\times{f}_1$\,mm  \cite{tokodi2017optimization}\\ \hline
   wavelength $\lambda$ & 1030\,nm& grating period d & 1/1500 $\text{mm}$ \\ \hline
    pulse duration $\tau_0$ (FWHM)  & 0.5\,ps \cite{fulop2012generation} & peak fluence& $10^8\sqrt{\tau_{0}}$\, mJ/cm$^2$ \cite{ravi2016pulse} \\ \hline
  phase-matched THz frequency  &0.3\,THz & absorption coefficient $\alpha(300\, \text{K},0.3 \,\text{THz})$ & 7/cm \cite{unferdorben2015measurement} 
 \\ \hline
    \end{tabular}
\end{center}
\end{table}

\begin{figure}[H]
\centering{
   \includegraphics[width=1\textwidth]{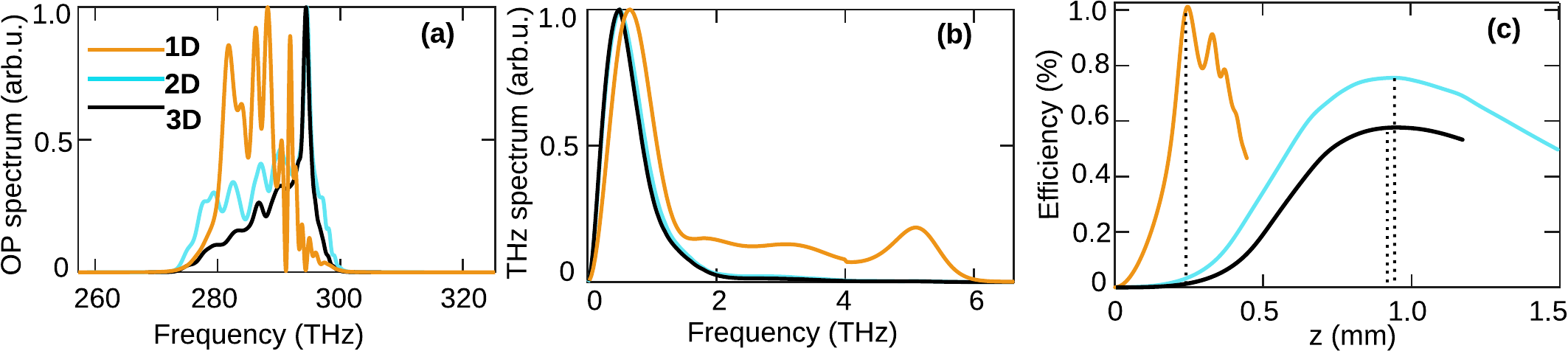}}
   \caption{Comparison of the results obtained from 1D+1, 2D+1 and 3D+1 simulations. (a), (b) and (c) are the output OP spectra, the output terahertz spectra and the efficiencies respectively. (a) and (b) are plotted at the location of maximum efficiency (marked by the dash lines in (c)).}\label{1-3compare}
\end{figure}

By comparing different models, one can see that the 1D+1 calculation neglects diffraction effects and, more importantly, the spatial walk-off between the terahertz and OP beams (the operator $2ik_{x0}\frac{\partial}{\partial x}$). The neglected terms are marked in Eqs. (\ref{eq1}) and (\ref{eq2}) by "neglected by 1D+1". Figure \ref{1-3compare} suggests that, by neglecting any spatial walk-off effect, the 1D+1 model overestimates the OP spectral broadening leading to a more pronounced stimulated Raman effect. Consequently, higher terahertz frequencies can be generated within the OP bandwidth via difference-frequency generation. This explains why the 1D+1 model predicts higher terahertz frequency content as compared to the 3D+1 model shown in Fig. 2(b). The 2D+1 model, though neglecting diffraction in the dimension $y$ (labeled by "neglected by 2D+1"), captures the broadening of the OP spectrum better than the 1D+1 case. Accordingly, the terahertz spectrum is predicted in very good agreement with the 3D+1 model. 

Figure \ref{1-3compare}(c) shows the computed conversion efficiency along the terahertz propagation direction ($z$). As can be seen, the 1D+1 model drastically underestimates the optimal interaction length, which also leads to a significant overestimation of the conversion efficiency. The interaction length is captured better by the 2D+1 model, while the conversion efficiency is still overestimated by about $25\%$ compared to the value obtained by a 3D+1 calculation, since the reduction of the OP fluence along the $y$ dimension is not accounted for in this model and only properly captured in the 3D+1 case.

It can be concluded that in order to capture the key characteristics of the OP and terahertz spectra, at least a 2D+1 model should be used while for a proper prediction of the conversion efficiency both 1D+1 and 2D+1 model overestimate the efficiency significantly. Here, a 3D+1 model is recommended. 

\section{Spatial dependence of the terahertz electric field}
Without loss of generality, the nonlinear interaction between the OP and the LN-crystal is numerically implemented in the $x-y-z$ coordinate frame where $x=0$ represents the apex location of the LN-crystal. Note that the OP beam size in the $x-y-z$ coordinates $\sigma_x$=$\sigma_{x'}/\cos(\gamma)$ is due to the projection onto the plane of the tilted pulse front. The simulations suggest that within an OP beam size range $\sigma_{y}=[0.5, 4.5]$\,mm (not shown), diffraction has a negligible effect on the terahertz generation process and the terahertz beam size scales as $\sigma_{y}/\sqrt{2}$. This agrees well with the analytic result in Eq. (\ref{polari}). In the following simulations, $\sigma_y$ is chosen to be $3.5$\,mm.

\begin{figure}[H]
\centering{
   \includegraphics[width=0.35\textwidth]{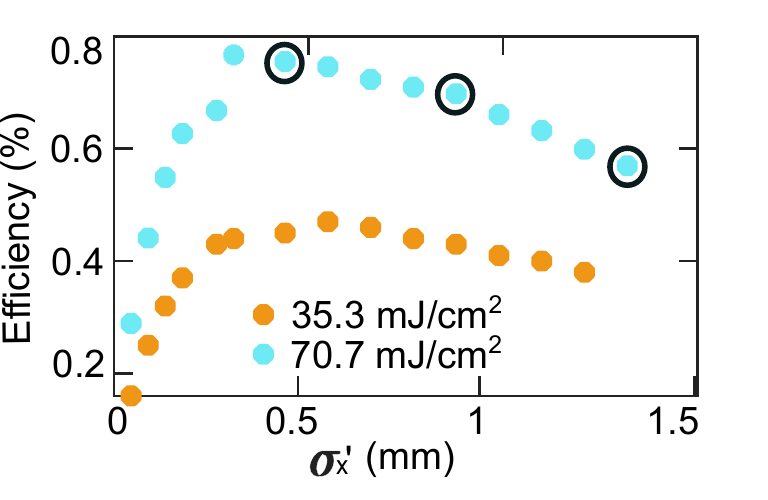}}
   \caption{With the input pump fluence  \text{70.7  mJ/cm$^2$} (blue dots) and \text{35.3  mJ/cm$^2$} (orange dots), the maximum terahertz generation efficiencies versus the OP beam size $\sigma_{x'}$, calculated by the 2D model, are presented. The black circles indicate 3 beam sizes chosen as examples in the following 3D+1 calculations.}\label{eff_bs}
\end{figure}

In Fig. \ref{eff_bs}, the maximum terahertz generation efficiency is plotted against the OP beam size $\sigma_{x'}$ for two different OP peak fluences. For this calculation, the 2D+1 model is chosen due to the high computational cost of the 3D+1 model. 

Due to the nature of the non-collinear phase-matching condition, the terahertz generation process requires different sections of the beam along the $x$ dimension to add up coherently in the emission direction $z$. In contrast with the OP beam size in the $y$ dimension, a small beam size along the $x$ dimension cannot produce high generation efficiencies due to the walk-off between the OP and the terahertz beam. On the other hand, if the beam size is too large, the terahertz radiation generated by the side of the OP at the farther side from the LN-crystal apex suffers from more absorption compared to the part closer to the apex. Thus, the generation efficiency shows a maximum as a function of OP beam size in the $x'$-direction. Additionally, since lower pump fluence leads to longer interaction length, the optimal pump beam size increases (see the orange dots in Fig. \ref{eff_bs}). 

As in experiments different OP beam sizes may be required in order to optimize the use of the available pump energy and limited crystal aperture, we select three pump sizes ($\sigma_{x'}=0.44\text{mm,}\,\,0.88 \text{mm and }1.32$ mm, marked by black circles in Fig. \ref{eff_bs}) for studying the spatio-temporal properties of the terahertz field using the 3D+1 model.

\begin{figure}[H]
\centering{
   \includegraphics[width=1\textwidth]{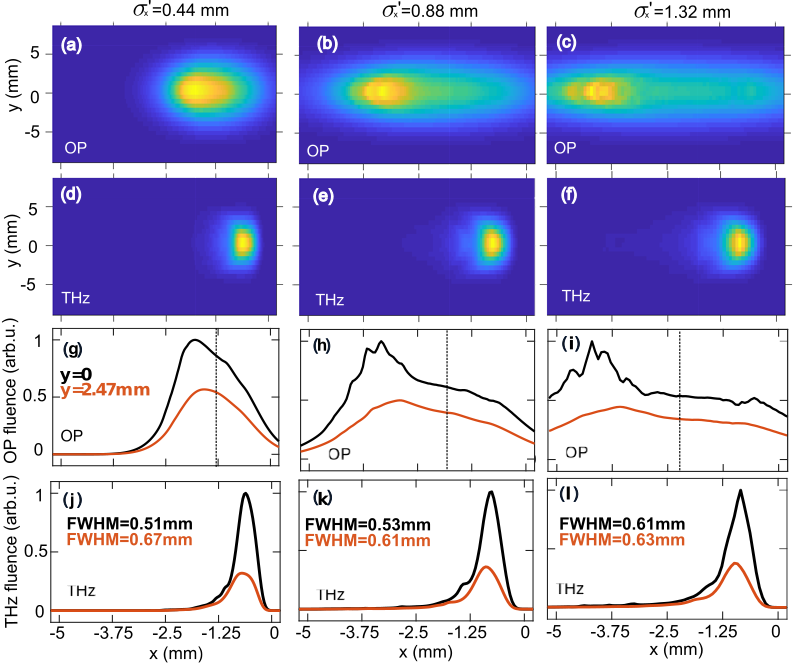}}
   \caption{Spatial dependence of the generated terahertz beams along $x$ and $y$ dimensions. (a-c) and (d-f) represent the OP and the terahertz beam profiles at the output surface of the LN-crystal respectively. (g-i) and (j-l) represent the OP and terahertz fluence respectively at a given position $y=0$ (black curve) and $y=\sigma_y/\sqrt{2}=2.47$ (red curve). The OP beam sizes at the input LN-crystal surface are $\sigma_x'=0.44\,\text{mm}\text{, }0.88\,\text{mm} \text{ and } 1.32\,\text{mm}$ in the $x'-y'-z'$ frame respectively. The center position of the OP beam is marked by the dashed line. The OP beam size in the $y$ dimension is chosen to be $\sigma_y=3.5\, \text{mm}$. The apex of the LN-crystal is located at $x=0$. }\label{beam_size_compare}
\end{figure}

In order to compare the terahertz fields generated by different spatial positions of the OP beam, the center of the OP (highest peak fluence $y=0$) and the side region of the OP (low fluence, $y=\sigma_y/\sqrt{2}=2.47$\,mm) within the optical beam are chosen. 
Figures \ref{beam_size_compare}(a)-\ref{beam_size_compare}(f) show the OP and the terahertz beam profiles at the output LN-crystal surface with different input OP beam sizes. For the beam sizes $\sigma_x'=0.44\,\text{mm}\,\,,0.88\,\text{mm}$ and $1.32\,\text{mm}$, the conversion efficiencies are 0.57$\%$, 0.54$\%$, 0.46$\%$ respectively. Figures \ref{beam_size_compare}(g)-\ref{beam_size_compare}(i) present the fluence of the OP at $y=0$ and $y=\sigma_y/\sqrt{2}$ with the center of the OP marked by the dashed lines. The fluence distribution indicates an energy shift towards the base of the LN-crystal. This shift is due to the non-collinear nature of the phase-matching mechanism, which causes the new optical frequencies of the OP, generated via the cascading effect, to propagate towards the base of the crystal. Surprisingly, the size of the OP does not strongly influence the size of the generated terahertz beams. Towards the base of the LN, the interaction length increases, the terahertz absorption and the nonlinear effect become more pronounced. The terahertz fields generated close to the base is absorbed before they reach the output LN surface. Thus, for large OP beam sizes, the OP farther from the LN apex is wasted. Figures \ref{beam_size_compare}(j)-\ref{beam_size_compare}(l) indicate that at a given OP beam size, a higher pump fluence leads to a smaller terahertz beam size compared with the case using a lower pump fluence. This finding agrees with the experimental results of C. Lombosi et al. \cite{lombosi2015nonlinear}. The terahertz beams are found to be symmetric along the y dimension.

\begin{figure}[H]
\centering{
   \includegraphics[width=1\textwidth]{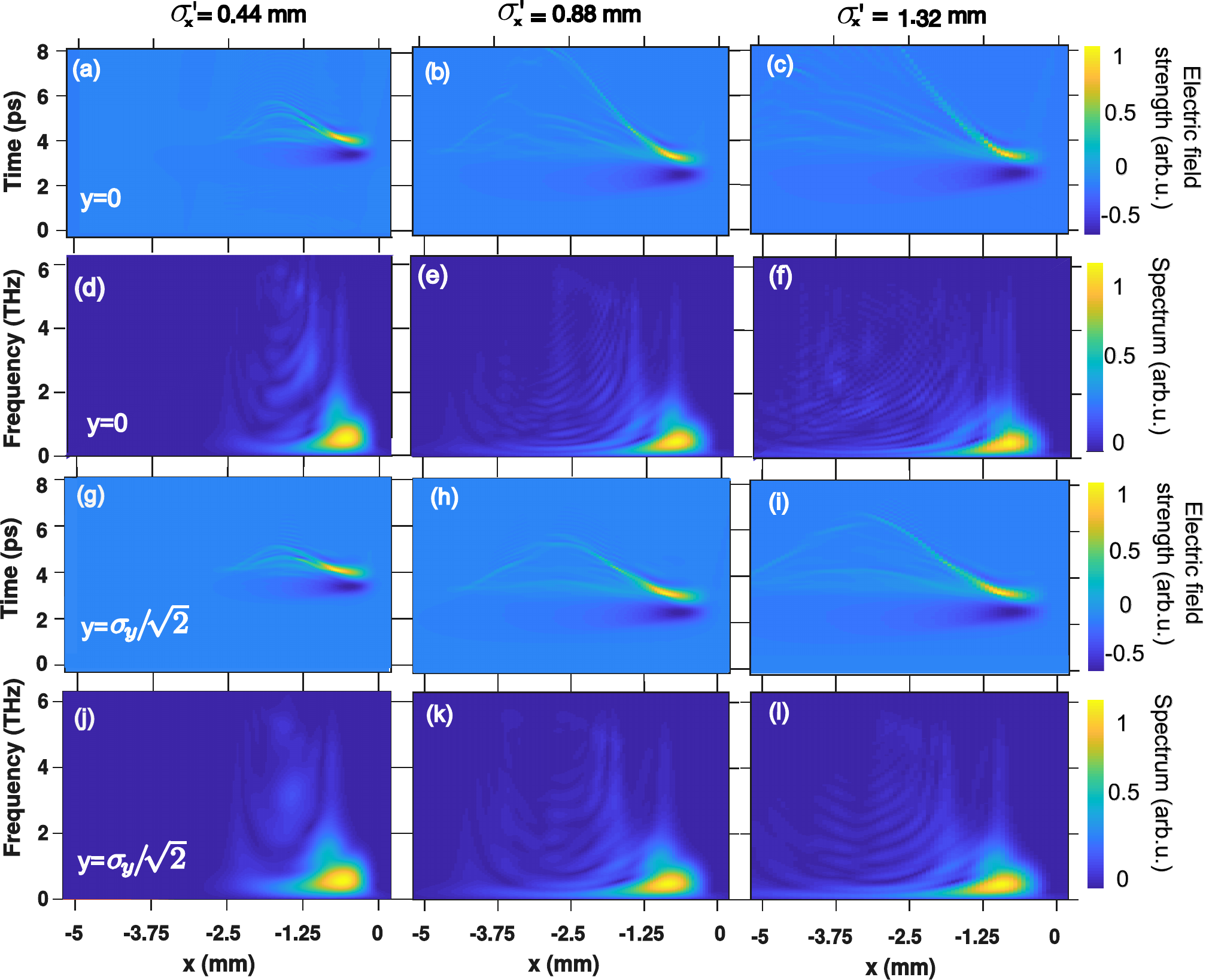}}
   \caption{Spatial dependence of the generated terahertz spectra and temporal profiles along the $x$ dimension. (a-c) and (d-f) are the terahertz electric field and the corresponding terahertz spectra with respect to $x$ at $y=0$.  (g-i) and (j-l) show the terahertz electric field and the corresponding terahertz spectra with respect to $x$ at $y=\sigma_y/\sqrt{2}$. \label{beam_size_compare_slice} }
\end{figure}

 Figures \ref{beam_size_compare_slice}(a)-\ref{beam_size_compare_slice}(c) and Figs. \ref{beam_size_compare_slice}(g)-\ref{beam_size_compare_slice}(i) show the spatially dependent electric fields at $y=\sigma_y/\sqrt{2}$ and $y=0$ respectively, with the corresponding terahertz spectra shown in Figs. \ref{beam_size_compare_slice}(d)-\ref{beam_size_compare_slice}(f) and Figs. \ref{beam_size_compare_slice}(j)-\ref{beam_size_compare_slice}(l). It can be seen that the few-cycle terahertz electric fields are only generated at the vicinity of the apex of the LN-crystal. This effect is seen to increase with larger pump beam sizes, as a larger fraction of the terahertz is generated farther from the apex and the terahertz electric fields deviate from a single-cycle waveform.  

\begin{figure}[H]
\centering{
   \includegraphics[width=0.35\textwidth]{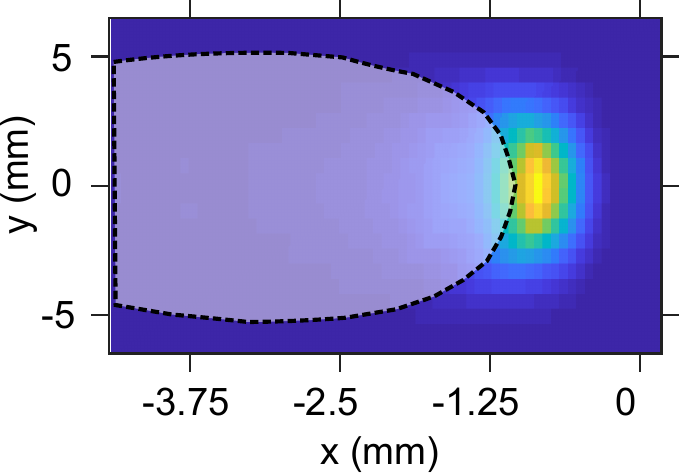}
   \caption{The example shown is for $\sigma_x'=1.32$\,mm, where the non-single-cycle region, $\Delta t(x,y)> 2\Delta t(x_p,y_p)$, is indicated by the grey region. The terahertz beam under the grey region contains up to 25\% of the total terahertz energy. }\label{e_field_quality}}
\end{figure}

To quantify the deviation from a single-cycle waveform, the root-mean-square pulse duration $\Delta t$ is chosen to evaluate the electric field distribution (see Eq.\ref{delta_t})). In Eq. (\ref{delta_t}), $\delta t=|t(x,y)-t(x,y)_p|$ , $t(x,y)_p$ is the time coordinate of the peak of the electric field at position $(x,y)$ and $I$ represents the intensity. The reference $\Delta t(x_p,y_p)$ is chosen at the position ($x_p,y_p$) where the terahertz peak fluence is located.

\begin{equation}\label{delta_t}
\Delta t(x,y)=\sqrt{\frac{\int[\delta t(x,y)]^2I(t,x,y)dt}{\int I(t,x,y)dt }-\left[\frac{\int\delta t(x,y)I(t,x,y)dt}{\int I(t,x,y)dt}\right]^2 }
\end{equation}  
The resulting map of $\Delta t(x,y)$ can be used to determine portion of the terahertz beam, where the electric field deviates significantly from a single-cycle format (see Fig. \ref{e_field_quality}). Figure \ref{e_field_quality} shows the terahertz beam generated by an OP with $\sigma_x=1.32$mm, where the part with $\Delta t(x,y)> 2\Delta t(x_p,y_p)$ is shaded grey to indicate the non-single-cycle content.

The grey region in Fig. \ref{e_field_quality} shows that high quality few-cycle pulses are only generated close to the apex of the crystal ($x = 0$). The terahertz beam profiles are symmetric along the $y$-direction. We find, that even though higher pump fluence ($y=0$) leads to a higher terahertz energy the portion of non-single-cycle content in the beam increases. It can be seen from Fig. \ref{beam_size_compare_slice} and Fig. \ref{e_field_quality} that the side of the OP beam (for example $y=\sigma_y/\sqrt{2}$\,mm) possess lower fluence, which leads to less OP spectral broadening in the terahertz generation process. Such a less broadened OP spectrum leads to a relatively homogeneous terahertz distribution along the $x$ dimension. For $\sigma_{x'}=$ 0.44\,mm,  0.88\,mm and 1.32\,mm  the $\Delta t(x,y)> 2\Delta t(x_p,y_p)$ region (grey region) take up 4\%, 20\% and 25\% of the total terahertz energy respectively. It can be seen that, with the increase of the OP beam size, the single-cycle region of the generated terahertz pulses reduces significantly. Additionally, due to the geometry of the nonlinear crystal, it is inevitable that the generated terahertz possesses a spatial inhomogeneity. In order to resolve this problem, a setup which combines a conventional tilted-pulse-front setup and a transmission stair-step echelon is promising to generate spatially homogeneous terahertz pulses \cite{palfalvi2017numerical}.

\section{Conclusion}
We have studied the spatial dependence of the terahertz electric field properties generated in a tilted-pulse-front setup. By comparing the 1D+1, 2D+1 and 3D+1 numerical models we found that the 1D+1 calculation cannot capture certain key features of the terahertz generation process. Additionally, the 3D+1 calculation shows that within the OP beam size range $\sigma_{y}=[0.5, 4.5]$ \,mm, diffraction in the dimension perpendicular to the pulse-front-tilt plane  does not have much influence on the terahertz generation process. For those cases, the 2D calculation (x-z coordinates) is a good approximation. 

Perpendicular to the pulse front tilt plane, in $y$-direction, the terahertz beam waist relates to the OP waist with  $\sigma_y/\sqrt{2}$, which is as expected for the second order nonlinear process. However, in the $x$ dimension, the OP beam size does not have a large impact on the  generated terahertz beam size. The terahertz pulses are generated close to the apex of the crystal and the single-cycle region is located close to the vicinity of the crystal apex. Thereby, large OP beam sizes lead to a reduced percentage of single-cycle content of the generated terahertz pulses. Attention must be paid to the fact that the terahertz generated farther from the apex of the crystal, which can possess a significant fraction of the total generated terahertz energy, suffers from poor electric field quality and temporal chirp. 

 In order to maximize the single-cycle content, the OP beam size along the $x$ dimension should be kept reasonably small, while the size along the $y$ dimension can be used to enlarge the pump beam if necessary. Another option is to reduce the OP input fluence. These findings are of particular relevance to carrier-envelope-phase sensitive terahertz applications and strong filed terahertz physics.

\section*{Funding}
European Union's Seventh Framework Program (FP7/2007-2013) through the Synergy Grant
AXSIS (609920);  The Hamburg Center for \text{Ultrafast} Imaging --Structure,
Dynamics and Control of Matter at the Atomic Scale (CUI, DFG-EXC1074).
\section*{Acknowledgment} 
We thank the
DESY HPC team for providing access and computation time on the DESY Maxwell cluster. Lu Wang would like to thank IMPRS (International Max Planck Research School for Ultrafast Imaging $\&$ Structural Dynamics) for the support both in science and life. 

\section*{Disclosures}
 The authors declare no conflicts of interest.

\bibliography{oe_3d.bib}

\end{document}